\documentclass[final]{aipproc}
\layoutstyle{6x9}

\def\ltsima{$\; \buildrel < \over \sim \;$}
\def\simlt{\lower.5ex\hbox{\ltsima}}
\def\gtsima{$\; \buildrel > \over \sim \;$}
\def\simgt{\lower.5ex\hbox{\gtsima}}

\begin{document}

\title{Extrasolar Planet Orbits and Eccentricities}

\author{Scott Tremaine}{address={Department of Astrophysical Sciences,
Princeton University, Princeton NJ 08544}} 

\author{Nadia L. Zakamska}{address={Department of Astrophysical Sciences,
Princeton University, Princeton NJ 08544}} 

\begin{abstract}

The known extrasolar planets exhibit many interesting and surprising
features---extremely short-period orbits, high-eccentricity orbits,
mean-motion and secular resonances, etc.---and have dramatically expanded our
appreciation of the diversity of possible planetary systems. In this review we
summarize the orbital properties of extrasolar planets. One of the most
remarkable features of extrasolar planets is their high eccentricities, far
larger than seen in the solar system. We review theoretical explanations for
large eccentricities and point out the successes and shortcomings of existing
theories.

\end{abstract}

\maketitle

Radial-velocity surveys have discovered $\sim 120$ extrasolar planets, and
have provided us with accurate estimates of orbital period $P$,
semimajor axis $a$, eccentricity $e$, and the combination $M\sin i$ of the
planetary mass $M$ and orbital inclination $i$ (assuming the stellar masses
are known). Understanding the properties and origin of the distribution of
masses and orbital elements of extrasolar planets is important for at least
two reasons.  First, these data constitute essentially everything we know
about extrasolar planets. Second, they contain several surprising features,
which are inconsistent with the notions of planet formation that we had
before extrasolar planets were discovered. The period distribution is
surprising, because no one predicted that giant planets could have orbital
periods less than 0.1\% of Jupiter's; the mass distribution is surprising
because no one predicted that planets could have masses as large as 10
Jupiters, or that there would be a sharp cutoff at this point; and the
eccentricity distribution is surprising because we believed that planets
forming from a disk would have nearly circular orbits.

\section{1. Mass and period distributions}

The mass distribution of extrasolar planets is sharply cut off above $10 M_J$,
where $M_J$ is the Jupiter mass \cite{zuck01,jor01}. The absence of companions
with masses in the range $10 M_J<M<100 M_J$ (the ``brown dwarf desert'')
provides the strongest reason to believe that the extrasolar planets are
formed by a different mechanism than low-mass companion stars, and thus the
criterion $M<10M_J$ is probably the least bad definition of an extrasolar
planet. The observed mass distribution for $M<10M_J$ can be modeled as a power
law $dn
\propto M^{-\alpha} dM$ with $\alpha=1.1\pm 0.1$
\cite{zuck01,hea99,step01,taba02}; in 
other words the distribution is approximately flat in $\log M$, at least over
the range $M_J\simlt M\simlt10M_J$. Weak evidence that this distribution can
be extrapolated to smaller masses comes from the solar system: fitting the
distribution of the masses of the 9 planets to a power law over the interval
from $M_{\rm Pluto}$ to $M_J$ yields $\alpha=1.0\pm 0.1$, with a high
Kolmogorov-Smirnov (KS) confidence level (over 99.9\%).

The period distribution can also be described by a power law to a first
approximation: $dn\propto P^{-\beta} dP$, where $\beta=0.73\pm 0.06$
\cite{taba02}. Some analyses \cite{hea99,step01,mz02} find a steeper
slope, $\beta \simeq 1$, probably because they do not correct for selection
effects (the reflex motion of the star due to a planet of a given mass
declines with increasing period). \citet{step01} and \citet{mz02} stress that
the distribution of periods (and eccentricities) is almost the same for
extrasolar planets and spectroscopic binaries---only the mass distributions
are different---and this striking coincidence demands explanation (see
\S2.7). Recall that the standard minimum solar nebula has surface density
$\Sigma(r)\propto r^{-1.5}$ \cite{hay81}. If the formation process assigned
this mass to equal-mass planets without migration we would have $\beta=0.67$,
consistent with the exponent found above for extrasolar planets.

\begin{figure}
\includegraphics[height=.35\textheight]{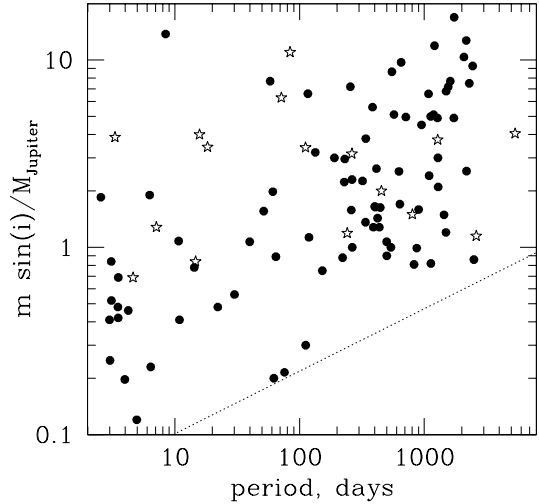}
\caption{Minimum masses versus periods of extrasolar planets. Planets found in
wide binary systems (from \cite{egge03}) are denoted by stars. The absence of
planets in the lower right part of the figure is due to the limited
sensitivity of the radial-velocity surveys (the dotted line shows an amplitude
of $10\hbox{\, m s}^{-1}$ for a planet on a circular orbit around a solar-mass
star). \label{pic_mass_period}}
\end{figure}

In Figure \ref{pic_mass_period} we plot the masses and periods of extrasolar
planets.\footnote{The data for figures are taken from
\url{http://www.obspm.fr/planets} as of October 2003.} Although power laws are
good first approximations to the mass and period distributions for $M<10 M_J$,
there have been a number of suggestions of additional structure. (i)
\citet{zuck02} and \citet{udry03} suggest that there are too few massive
planets ($M\simgt3M_J$) on short-period orbits ($P\simlt50$ days), although
(ii) Zucker \& Mazeh argue that this shortfall is not present if the host star
is a member of a wide binary system (denoted by star symbols in Figure
\ref{pic_mass_period}). In addition, \citet{udry03} suggest that there is
(iii) a deficit of planets in the range $10\hbox{ d}\simlt P\simlt 100\hbox{
d}$; and (iv) a deficit of low-mass planets ($M\simlt0.8 M_J$) on long-period
orbits ($P\simgt 100\hbox{ d}$). We do not find the latter two features
persuasive, since their significance has not yet been verified by statistical
tests, and of course low-mass planets on long-period orbits produce the
smallest reflex velocity and so are most difficult to detect.

Whether or not these features exist, in our view the most impressive feature
of the period distribution is that it is described so well by a power law. In
particular, the mechanisms that might halt planetary migration often operate
most effectively close to the star \cite{ter03}, so it is natural to expect a
``pile-up'' of planets at orbital periods of a few days; however, there is
only weak evidence for an excess of this nature compared to a power-law
distribution.

\section{2. Eccentricity distribution}

One of the remarkable features of extrasolar planets is their large
eccentricities: the {\em median} eccentricity of 0.28 is larger than the {\em
maximum} eccentricity of any planet in the solar system. The eccentricity
distribution of extrasolar planets differs from that of solar system planets
at the 99.9\% KS confidence level.

Observational selection effects in radial-velocity surveys could either favor
or disfavor detection of high-eccentricity planets: as the eccentricity grows
at fixed period the periastron velocity grows, but the time spent
near periastron declines. The limited simulations that have been done
\cite{fis92,maz96} suggest that these two effects largely cancel, so the
eccentricity-dependent selection effects are small. Thus, the observed
distribution should accurately reflect the eccentricities of massive,
short-period extrasolar planets.

Planets that form from a quiescent disk are expected to have nearly circular
orbits, and thus the high eccentricities of extrasolar planets demand
explanation. A wide variety of eccentricity excitation mechanisms has been
suggested so far.

\subsection{2.1. Interactions with the protoplanetary gas disk}

Gravitational interactions between a planet and the surrounding protoplanetary
disk can excite or damp the planet's eccentricity. These interactions are
concentrated at discrete resonances, the most important of which are given by
the relation \cite{gold78,gold80,gold81,ward86,arty93}
\begin{equation}
\Omega + {\epsilon\over m}(\Omega-\dot\varpi)= \Omega_p + {\epsilon_p\over
m}(\Omega_p-\dot\varpi_p);
\label{eq:res}
\end{equation}
here $\Omega$ is the mean angular speed, $\dot\varpi$ is the apsidal
precession rate, the integer $m>0$ is the azimuthal wavenumber,
$\epsilon=0,\pm1$, and the unsubscripted variables refer to the disk while
those with the subscript ``p'' refer to the planet. Resonances with
$\epsilon=0$ are called corotation resonances and those with $\epsilon=+1$ and
$-1$ are called outer and inner Lindblad resonances, respectively. Resonances
with $\epsilon=\epsilon_p$ are called ``coorbital'' resonances since the
resonant condition (\ref{eq:res}) is satisfied when the gas orbits at the same
angular speed as the planet; coorbital resonances are absent if the planet
opens a gap in the disk. Resonances that are not coorbital are called external
resonances. Apsidal or secular resonances have $m=1$, $\epsilon=\epsilon_p=-1$
so the resonant condition is simply $\dot\varpi=\dot\varpi_p$; in this case
the collective effects in the disk (mostly pressure) only have to compete with
differential precession rather than with differential rotation, so the
resonance is much broader.

Interactions at external Lindblad and corotation resonances
excite and damp the planet's eccentricity, respectively \cite{gold80}. In the
absence of other effects, damping exceeds driving by a small margin ($\sim
5\%$). However, corotation resonances are easier to saturate, by trapping
the orbital angular speed into libration around the pattern speed; if the
corotation resonances are saturated and the Lindblad resonances are not, then
the eccentricity can grow.

Interactions at coorbital resonances damp the planet's eccentricity
\cite{ward86,arty93}, but these only operate in a gap-free disk. Apsidal
resonances can also damp eccentricity \cite{ward98,ward00}, but at a rate that
depends sensitively on the disk properties \cite{gold03}.

Given these complexities, about all we can conclude is that either
eccentricity growth or damping may occur, depending on the properties of the
planet and the disk. Numerical simulations have so far provided only limited
insight: two-dimensional simulations \cite{nels00,papa01a} show eccentricity
growth if and only if the planet mass exceeds 10--$20 M_J$, but the relation
of these results to the resonant behavior described above remains unclear.

\subsection{2.2. Close encounters between planets}

In the final stages of planet formation dynamical instability can develop,
either as the masses of the planets increase due to accretion or as their
orbital separation decreases due to differential migration. The instability
usually leads one or more planets to be ejected from the system or to collide
with one another or with the central star (this last outcome is relatively
rare). The surviving planets are left on eccentric orbits
\cite{rasi96,weid96,lin97,ford01}. 

A great attraction of this mechanism is that it makes calculable predictions;
unfortunately, the predictions have some difficulty matching the observations:
(i) It has been suggested that this mechanism could produce the `hot
Jupiters'---planets such as 51 Peg B that are found on low-eccentricity, very
short-period orbits (a few days) \cite{rasi96}---by close encounters of
planets at distances of a few AU, which throw one planet onto a highly
eccentric orbit that is later circularized by tidal dissipation. However, the
frequency of hot Jupiters is far larger than this mechanism could produce
\cite{ford01}. (ii) The median eccentricity of the surviving planet following
an ejection is about 0.6, compared to 0.3 in the observed systems
\cite{ford01}; on the other hand, so far the simulations have focused on
equal-mass planets and the eccentricity is likely to be lower when a more
massive planet ejects a less massive one. (iii) Collisions between planets
lead to a population of collision remnants on nearly circular orbits, which is
not observed \cite{gold03,ford01}. To avoid collisions would require that the
planets are much more compact than Jupiter (i.e. the escape speed from the
planet must be much larger than the orbital speed).

\begin{figure}
\includegraphics[height=.35\textheight]{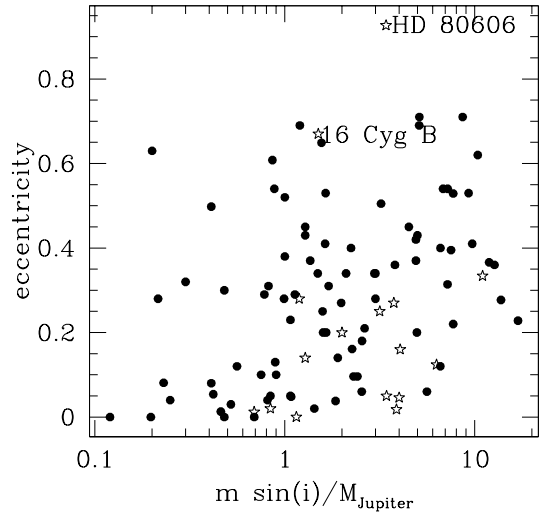}
\caption{Eccentricity versus minimum mass for extrasolar planets. Planets
found in wide binaries (from \cite{egge03}) are shown with stars, and two
examples with extreme eccentricities (16 Cyg B and HD 80606) are
labeled.  \label{pic_ecc_mass}} 
\end{figure}

Another potential problem is the predicted mass-eccentricity relation: in this
scattering process, low-mass planets are naturally expected to be excited to
higher eccentricities. Unfortunately, so far no simulations have looked for
mass-eccentricity correlations in a large sample of simulated planetary
systems, so we do not know how strong the correlation should be. The
eccentricities and masses of the known extrasolar planets show a weak
correlation (at about 90\% confidence level) in the {\em opposite} sense: more
massive planets seem to have higher eccentricities (Figure
\ref{pic_ecc_mass}).

A related possibility is that the eccentricity is excited by interactions with
massive planetesimals \cite{murr98,murr02}. Normally, close encounters with
small bodies tend to damp the eccentricity and resonant interactions excite
eccentricity. Because this process relies on interactions with solid bodies,
rather than gas giant planets, it requires a rather massive planetesimal disk,
comparable to the mass of the planet, which in turn requires a gaseous
protoplanetary disk that is even more massive, $\simgt 0.1M_\odot$.

\subsection{2.3. Resonant interactions between planets}

Many satellites in the solar system, as well as Neptune and Pluto, are locked
in orbital resonances. The formation and evolution of these resonances has
been thoroughly studied (see for example \cite{pea99} and references therein).

Typically, resonance capture in the solar system occurs because of convergent
outward migration, in which an inner satellite migrates outward faster than an
outer satellite, thereby entering a mean-motion resonance. Continued migration
after resonance capture excites the eccentricities of the resonant bodies; for
example, this process is believed to be responsible for Pluto's eccentricity
of 0.25 \cite{malh93}.

In contrast, the known extrasolar planets have presumably migrated {\em
inward}. In this case, migration after resonance capture excites
eccentricities even more efficiently. For example, a test particle that is
captured into the 2:1 resonance with an exterior planet will be driven to an
eccentricity of unity when the planet has migrated inward by a factor of four
following capture \cite{yu01}. \citet{lee01} have
analyzed the dynamics of the planets in GJ 876, which are locked in a 2:1
resonance; they argue that to avoid exciting the eccentricities above the
observed values requires either (i) extremely strong eccentricity damping, or
(ii) fine-tuning the resonance capture to occur just before migration stops. 

\citet{chia02} point out that eccentricities can also be excited by divergent
migration, in which the ratio of the semimajor axes of two planets increases;
the planets traverse a series of resonances, each of which excites additional 
eccentricity.

There are at least two concerns with models based on resonant interactions:
(i) this mechanism requires at least two planets, whereas only a single planet
has been discovered so far in most of the known extrasolar planetary systems;
(ii) like close encounters, resonant interactions are expected to excite
low-mass planets to higher eccentricities, but the observed mass-eccentricity
correlation is small, and if anything in the opposite sense (Figure
\ref{pic_ecc_mass}). 

\subsection{2.4. Secular interactions with a distant companion star}

A planet's eccentricity can be excited by secular interactions with a distant
companion that does not lie in the planet's orbital plane---this is sometimes
called the Kozai mechanism
\cite{koza62,holm97,inna97,maze97,krym99,ford00}. This mechanism has the
interesting property that the separation and mass of the companion affect the
period but not the amplitude of the eccentricity oscillation; thus even a weak
tidal force from a distant stellar or planetary companion can excite a large
eccentricity. However, the precession rate due to the companion must exceed
the precession rate due to all other effects: other planets in coplanar
orbits, any residual planetesimal disk, general relativity, etc.

There are several straightforward predictions for this mechanism: (i) There
should be no correlation between the planet's mass and eccentricity, which is
marginally consistent with the weak correlation seen in Figure
\ref{pic_ecc_mass}. (ii) High-eccentricity planets should be found in binary
star systems. This does not appear to be the case (Figure \ref{pic_ecc_mass}),
although two of the highest eccentricities, 16 Cyg B and HD 80686, are found
in binary systems. Perhaps some of the other high eccentricities are excited by
unseen companions---planets or brown dwarfs. (iii)
Coplanar multi-planet systems should have low eccentricities, since their
mutually induced precession is far larger than the precession due to a distant
companion, thereby suppressing the Kozai mechanism. This prediction is not
supported by the dozen or so multi-planet systems, which have substantial
eccentricities, although in most cases we have no direct evidence that these
systems are coplanar (see \S3).

\subsection{2.5. Phenomenological diffusion}

Given the limited success of the models we have described so far, it is useful
to ask what constraints we can place on the eccentricity excitation mechanism
from simple phenomenological models. Here we explore the possibility that the
excitation can be modeled as a diffusion process in phase space, with an
eccentricity-dependent diffusion coefficient $D(e)$.  

We shall work with the eccentricity vector ${\bf
e}=(e\cos\varpi,e\sin\varpi)$, since its components are approximately
canonical variables for small eccentricities.  Let $n({\bf e})d{\bf
e}$ be the number of planets with eccentricity vector in the range $d{\bf
e}$. Then the diffusion equation in the eccentricity plane can be written as
\begin{equation}
\frac{\partial n(e,t)}{\partial t}=\nabla(D\nabla
n)=\frac{1}{e}\frac{\partial}{\partial e}\left[eD(e)\frac{\partial
n(e,t)}{\partial e}\right]. 
\end{equation}
We solve this equation for two cases: (i) a single initial burst of formation
at $t=0$ ($n(e,0)\propto\delta(e)/e$); (ii) continuous formation at a constant
rate ($n(0,t)=$const). Since planets are ejected from the system when they
reach $e=1$, there is also a boundary condition $n(1,t)=0$. We assume that the
diffusion coefficient is a power law, $D(e)\propto e^p$.

\begin{figure}
\includegraphics[height=.4\textheight]{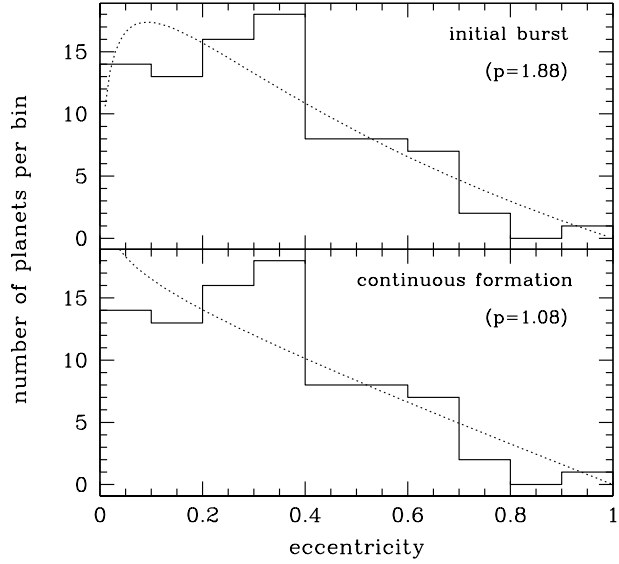}
\caption{Eccentricity distribution: observed (histogram) and predicted by the
diffusion models (dotted lines). Tidally circularized planets ($P<6$ days) are
excluded from the analysis. The diffusion coefficient in the eccentricity
vector plane is $D(e)\propto e^p$.
\label{pic_ecc_dist}}  
\end{figure}

The best-fit ccentricity distributions that emerge from the diffusion models
are plotted in Figure \ref{pic_ecc_dist}. They are shown at late times when
the planets in the initial-burst model have diffused away from the center of
the eccentricity plane ($e=0$) and the continuous-formation model has reached
its steady state $n(e)\propto e^{1-p}-e$. Both the initial-burst model and the
continuous-formation model produce satisfactory fits (with KS confidence level
about 60\%), so we cannot distinguish between the two types of models by
comparing to the current observations; however, the predicted behavior is very
different near $e=0$ so the models should be distinguishable with more data.

In both cases, the best-fit values of $p$ ($p=1.9$ for the initial-burst
model and $p=1.1$ for the continuous-formation model) suggest that
eccentricity diffusion is produced by a mechanism that is ineffective for
circular orbits. 

\subsection{2.6. Propagation of eccentricity disturbances from other planets}

Stars in the solar neighborhood approach passing stars to within about 400 AU
during a lifetime of $10^{10}$ y. Such encounters cannot excite the
eccentricity of a single planet on a orbit like those of the known extrasolar
or solar system planets. However, disks of gas and dust around young stars are
observed to extend to radii of hundreds of AU, and it is plausible to assume
that these disks form planetesimals or planets that survive to the present at
radii $\sim 10^2$ AU. If so, then passing stars can efficiently excite the
eccentricities of the outer bodies in the system, and the eccentricity
disturbance can then propagate inward via secular interactions between the
planets or planetesimals, much like a wave. This process is described more
fully in the article ``Propagation of eccentricity disturbances in planetary
systems'' \cite{zaka03} elsewhere in these proceedings. The most important
disk properties that determine the efficiency of this excitation mechanism are
the radial extent of the disk and the steepness of its smeared-out
surface-density profile; the total mass of the disk is not a factor. If the
surface density is a power law, $\Sigma(r)\propto r^{-q}$, efficient
eccentricity excitation at small radii requires a rather flat profile,
$q\approx 1$, compared to the minimum solar nebula profile $q\approx 1.5$.

\subsection{2.7. Formation from a collapsing protostellar cloud}

It is conventional to assume that the components of binary star systems form
simultaneously from condensations in a collapsing protostellar cloud (and
hence have high-eccentricity orbits), whereas planets form later from the
protoplanetary disk (and hence initially have low-eccentricity orbits). The
two populations of companions seem to be clearly separated by the brown-dwarf
desert.

Thus it is remarkable that the eccentricity and period distributions of
extrasolar planets and low-mass secondaries of spectroscopic binaries
are almost indistinguishable \cite{zuck01,step01}. This observation
suggests that perhaps the extrasolar planets, like binary stars,
formed in the collapsing protostellar cloud \cite{black97,papa01b,terq02}.

Simulations by \citet{papa01b} suggest that this process preferentially forms
either ``hot Jupiters'' or planets with high eccentricities and large
semimajor axes; although Papaloizou \& Terquem did not carry out a detailed
comparison, it seems likely that, with some tuning, their simulations could
reproduce the distributions of periods and eccentricities of the known
extrasolar planets. A possible concern is that the lower limit for
opacity-limited fragmentation is a few Jupiter masses, and many planets are
known to have smaller masses. \citet{terq02} postulate that the low-mass
planets formed in a disk and their eccentricities were later excited by
encounters with high-mass planets formed by fragmentation.

\section{3. Inclination Distribution}

So far there is little direct evidence that the planets in multi-planet
systems have small mutual inclinations, as we might expect if they formed from
a disk. \citet{chia01} argue that if the apsidal alignment between $\upsilon$
And C and D ($\Delta\omega=5^\circ\pm5^\circ$) is not a coincidence, then
these two planets must have mutual inclination $\simlt
20^\circ$. \citet{laug02} conclude that stability considerations restrict the
mutual inclination between the two planets in 47 UMa to $\simlt 40^\circ$.

In multi-planet systems with large mutual inclinations the Kozai mechanism can
lead to large inclination and eccentricity oscillations, which promote
instability by enabling close encounters between planets or with the central
star. Thus, even if planetary systems are formed from collapsing clouds, as
suggested in \S2.7, multi-planet systems may be restricted to a
disk-like configuration, containing planets on both prograde and retrograde
orbits. 

\section{4. Conclusions}

Planetary eccentricities can be excited by a variety of mechanisms, and it is
likely that more than one of these plays a role in determining the
eccentricity distribution of extrasolar planets. As the planet search programs
continue to find planets of lower and lower masses, and longer and longer
periods, which resemble more closely the giant planets in the solar system,
the issue of why our planets have much lower eccentricities than their
extrasolar analogs becomes more acute. 

\bibliographystyle{aipproc}

\end{document}